\documentclass[12pt]{article}
\usepackage{amsmath,amssymb}
\usepackage[english]{babel}
\usepackage[cp866]{inputenc}

\usepackage{hyperref}

\numberwithin{equation}{section}

\newcommand{\bq}{\begin{eqnarray}}
\newcommand{\eq}{\end{eqnarray}}
\newcommand{\bbq}{\begin{equation*}}
\newcommand{\eeq}{\end{equation*}}

\newcommand{\ra}{\rightarrow}
\newcommand{\la}{\Lambda_Q}
\newcommand{\mph}{\mu_{\Phi}}
\newcommand{\ov}{\overline}
\newcommand{\lym}{\Lambda_{SYM}}

\newcommand{\bo}{{\rm b_o}}
\newcommand{\nd}{{\ov N}_c}

\newcommand{\no}{n_1}
\newcommand{\nt}{n_2}

\newcommand{\bd}{{\rm\ov b}_{\rm o}}

\newcommand{\qq}{{\ov Q}Q}

\newcommand{\mo}{\mu_{\Phi, 0}}
\newcommand{\mI}{\mu_{\Phi, 1}}

\newcommand{\qtp}{\mu^{\rm pole}_{q,2}}

\newcommand{\muo}{{\ov\mu}_{{\rm gl},\,1}}

\newcommand{\mgt}{\mu_{\rm gl, 2}}
\newcommand{\mgo}{\mu_{\rm gl, 1}}
\newcommand{\dq}{\textsf{q}}
\newcommand{\odq}{\ov{\textsf{q}}}

\newcommand{\Qo}{({\ov Q}Q)_1}
\newcommand{\Qt}{({\ov Q}Q)_2}
\newcommand{\cw}{{\mathcal W}}
\newcommand{\qo}{({\ov q}q)_1}
\newcommand{\qt}{{(\ov q}q)_2}

\begin{document}

\begin{center}{\bf \large Mass spectra in $\mathbf{{\cal N}=1\,\, SQCD}$  with additional colorless fields\\
and problems with Seiberg's duality} \end{center}

\vspace*{-2mm}
{\hspace*{7.5cm} \bf Victor L. Chernyak }\\
\vspace*{-2mm}

\hspace*{1cm}  Budker Institute of Nuclear Physics and Novosibirsk State University, Novosibirsk, Russia \\
\vspace*{-2mm}

{\hspace*{1.2cm} Talk given at Session of Nucl. Phys. section of RAS, Novosibirsk, Russia, 10-12 March\\

\vspace*{-2mm}
{\hspace*{8.5cm} {\bf Abstract}}

{\hspace*{2mm} Considered is the direct ${\cal N}=1$ SQCD-like $\Phi$-theory with $SU(N_c)$ colors and $3N_c/2< N_F<2N_c$ flavors of light quarks ${\ov Q},\,Q$. Besides, it includes $N^2_F$ additional colorless but flavored fields $\Phi_{i}^{j}$ with the large mass parameter $\mph\gg\la$, interacting with quarks through the Yukawa coupling. In parallel, is considered its Seiberg's dual variant, i.e. the $d\Phi$-theory with $(N_F-N_c)$ dual colors, $N_F$ flavors of dual quarks ${\ov q},\,{q}$ and $N_F^2$ elementary mion fields $M^i_j\ra ({\ov Q}_j Q^i)$.

In  considered here vacua, the quarks of both theories are in the conformal regimes at scales $\mu<\la$. The mass spectra are calculated in sections 4 and 5. It is shown that {\bf they are different in the direct and dual theories}, in disagreement with the Seiberg hypothesis about equivalence of two such theories.

Besides it is shown in the direct theory that a qualitatively new phenomenon takes place: the seemingly heavy fields $\Phi$ `{\bf return back}' and there appear {\bf two additional generations of light $\Phi$-particles} with small masses $\mu^{\rm pole}_{2,3}(\Phi)\ll\la$.

In Conclusions also presented comparison of mass spectra of these two theories for such values of parameters when the direct theory is in the very strong coupling regime, while the dual one is in the weak coupling IR-free logarithmic regime. It is shown that mass spectra of these two theories are {\bf parametrically different} in this case.

\tableofcontents
\numberwithin{equation}{section}

\newpage
\section{Definitions and some generalities}

\quad {\large\bf Direct $\mathbf \Phi$ - theory}\\

\hspace*{4mm} The field content of this direct ${\cal N}=1\,\,\,\Phi$-theory includes $SU(N_c)$ gluons and $3N_c/2< N_F<2N_c$ flavors of light quarks ${\ov Q}_j, Q^i$. Besides, there are $N^2_F$ colorless but flavored fields $\Phi^{j}_{i}$ (fions) with the large mass parameter $\mph\gg\la$.

The Lagrangian at the scale $\mu=\la$ in superfield notations looks as (the exponents with gluons in the Kahler term are implied here and everywhere below, $\nd=N_F-N_c$):
\bq
L=\int d^4 x\int d^2{\ov\theta}d^2\theta\, K(x,{\ov\theta},\theta)\,+ \Bigl (\cw+h.c.\Bigr ),\quad
\cw=\int d^4 x\int d^2\theta\, \cw_{tot}(x,\theta)\,, \label{(1.1)}
\eq
\bbq
K={\rm Tr}\,\Bigl (\Phi^\dagger \Phi\Bigr )+{\rm Tr}\Bigl (\,Q^\dagger Q+(Q\ra {\ov Q})\,\Bigr )\,,\quad \cw_{\rm tot}=\cw_{\rm gauge}+\cw_{\rm matter}\,,\quad \cw_{\rm gauge}=-\frac{2\pi}{\alpha(\mu,\la)} S\,,
\eeq
\vspace*{-3mm}
\bbq
\cw_{\rm matter}=\cw_Q+\cw_{\Phi}\,,\quad \cw_Q={\rm Tr}\,{\ov Q} m^{\rm tot}_Q Q ={\rm Tr}\,{\ov Q}(m_Q-\Phi) Q, \quad \cw_{\Phi}=\frac{\mph}{2}\Biggl [{\rm Tr}\,(\Phi^2)-\frac{1}{\nd}\Bigl ({\rm Tr}\,\Phi\Bigr )^2\Biggr ]\,.
\eeq
Here\,: $\mph$ and $m_Q$  are the mass parameters, $S=W^{A}_{\beta}W^{A,\,\beta}/32\pi^2$, where $W^A_{\beta}$ is the field strength of the gauge superfield, $A=1...N_c^2-1,\, \beta=1,2$,\, $\alpha(\mu,\la)=g^2(\mu,\la)/4\pi$ is the gauge coupling with its scale factor $\la$. Except for section 3, this normalization of fields is used everywhere below in the text.

Therefore, finally, the $\Phi$-theory we deal with has the parameters\,: $N_c,\,N_F,\,\mph,\,\la,\, m_Q$, with the {\bf strong hierarchies} $\mph\gg\la\gg m_Q$. Everywhere below in the text the mass parameter $\mph$ is in the range: $\la\ll\mph\ll\mo=\la (\la/m_Q)^{(2N_c-N_F)/N_c}$.
\footnote{\,
Here and below: $A\approx B$ means equality up to small corrections, $A\gg B$ has to be understood as $|A|\gg |B|$,\, $A\sim B$ means the same power dependence of A and B on small parameters $m_Q/\la\ll 1$,\, $\la/\mph\ll 1$ and $Z_q\ll 1$, up to a constant factor.
}

\vspace*{0.3cm}
\quad {\large \bf Dual $\mathbf d\Phi$-theory}\\

\vspace*{-1.9mm}
In parallel with the direct $\Phi$-theory with $3N_c/2<N_F<2N_c$\,,\, we consider also the Seiberg dual variant \cite{S2} (the $d\Phi$-theory). The dual Lagrangian at $\mu=\la$ looks as
\bbq
{\ov K}={\rm Tr}\,\Bigl (\Phi^\dagger \Phi\Bigr )+ {\rm Tr}\Bigl ( q^\dagger q + (q\ra\ov q)\, \Bigr )+{\rm Tr}\, \frac{M^{\dagger} M}{f^2 Z^2_q\la^2}\,, \quad {\ov\cw}={\ov\cw}_{\rm gauge}+{\ov\cw}_{\rm matter},\quad {\ov\cw}_{\rm gauge}=\, -\,\frac{2\pi}{\ov \alpha(\mu=\la)}\, {\ov S}\,,
\eeq
\vspace*{-4mm}
\bq
{\ov\cw}_{\rm matter}=\cw_{\Phi}+{\cw}_{M\Phi}+\cw_q\,,\quad{\cw}_{M\Phi}={\rm Tr}\, M(m_Q-\Phi),\quad {\cw}_q= -\frac{1}{Z_q\la}\,\rm {Tr}
\Bigl ({\ov q}\,M\, q \Bigr )\,.\label{(1.2)}
\eq

Here\,:\, the number of dual colors is $\nd=(N_F-N_c)$ and $M^i_j\ra ({\ov Q}_j Q^i)$ are $N_F^2$ elementary mion fields, ${\ov a}(\mu)=\nd{\ov g}^2(\mu)/8\pi$ is the dual running gauge coupling (with its scale parameter $|\Lambda_q|=\la$),\,\,${\ov S}={\rm \ov W}^{B}_{\beta}{\rm \ov W}^{B,\,\beta}/32\pi^2,\,\,B=1...(\nd^2-1)$,\,\, ${\rm \ov W}^B_{\beta}$ is the dual gluon field strength. The factors $a_f=\nd f^2/2\pi$ and $Z_q$ in \eqref{(1.2)} are $O(1)$ at $\bd/N_F=O(1)$ (and are omitted below in this case), but are parametrically small at $\bd/N_F \ll 1\,:\, a_f=O(\bd/N_F)$, while
$Z_q$ is exponentially small. (And $Z_q$ is accounted for then, see Conclusions).

At $3/2<N_F/N_c<2$ this dual theory can be taken as UV free at $\mu\gg\la$. We consider it below at $\mu\leq\la$ only where, according to Seiberg's hypothesis, it becomes equivalent to the direct $\Phi$-theory.

Really, {\bf all $N_F^2$ fields $\Phi^j_i$ remain always too heavy and dynamically irrelevant in this $d\Phi$-theory} at $3 N_c/2<N_F<2 N_c$ and $\mu<\la$, so that they can be integrated out once and forever and, finally, we write the Lagrangian of the dual theory at $\mu=\la$ in the form
\bbq
K= {\rm Tr}\Bigl ( q^\dagger q +(q\ra\ov q) \Bigr )+{\rm Tr}\,\frac{M^{\dagger}M}{f^2 Z^2_q\la^2}\,,\quad
{\ov\cw}_{\rm matter}=\cw_M+\cw_q\,,
\eeq
\vspace*{-4mm}
\bq
\cw_M=m_Q{\rm Tr}\,M -\frac{1}{2\mph}\Biggl [{\rm Tr}\, (M^2)- \frac{1}{N_c}({\rm Tr}\, M)^2 \Biggr ]\,,\quad\cw_q= -\,\frac{1}{Z_q\la}\,\rm {Tr} \Bigl ({\ov q}\,M\, q \Bigr )\,.\label{(1.3)}
\eq

The gluino condensates of the direct and dual theories are matched in all vacua, $\langle{-\,\ov S}\rangle=\langle S\rangle=\langle\lym\rangle^3$, as well as $\langle M_j^i(\mu=\la)\rangle=\langle M_j^i\rangle=\langle{\ov Q}_j Q^i (\mu=\la)\rangle=\langle{\ov Q}_j Q^i\rangle,\,\, \langle{\ov Q}_j Q^i\rangle=\sum_{a=1}^{N_c}\langle{\ov Q}_j^a Q^i_a\rangle$.\\

Besides, the perturbative NSVZ $\beta$-function \cite{NSVZa} for (effectively) massless ${\cal N}=1$ SQCD is used
\bq
\frac{d}{d \ln\mu}\,\frac{1}{a(\mu)}=\beta(a)=\frac{1}{1-a(\mu)}\Bigl [\frac{\bo}{N_c}-\frac{N_F}{N_c}\gamma_Q(a)\Bigr ]\,,\quad a(\mu)=\frac{N_c g^2(\mu)}{8\pi^2}\,,\quad \bo=3N_c-N_F\,, \label{(1.4)}
\eq
where $\gamma_Q$ is the quark anomalous dimension (and similarly in the dual theory: $N_c\ra\, \nd,\,\, \gamma_Q\ra\, \gamma_q,\,\,\, a=N_c g^2/8\pi^2\ra\, {\ov a}=\nd{\ov g}^2/8\pi^2,\, a_f=\nd f^2/2\pi,\, \bo=(3 N_c-N_F)\ra\, \bd=(3\nd-N_F)\,)$.

We take below (except for Conclusions): $\bo/N_F$ and $\bd/N_F$ as $O(1)$. Then $Z_q$ and $a_f$ are both $O(1)$ and are omitted.

Because the range $3N_c/2<N_F<2N_c$ considered here is within the conformal window $3N_c/2<N_F<3 N_c$, both the direct and dual theories (which are in the logarithmically weak UV free regime at $\mu\gg\la$) {\bf enter smoothly the conformal regime} at $\mu<\la$, with frozen couplings:\,$a(\mu<\la)=a_{*}=O(1),\,\, {\ov a}(\mu<\la)={\ov a}_{*}=O(1),\, a_f(\mu<\la)=a^{*}_f=O(1)$\, (until this conformal regime with effectively massless quarks, gluons and mions $M^i_j$ is broken by particles masses at lower energies). Then, the anomalous dimensions of all fields and so the corresponding renormalization factors of all Kahler terms are known in the conformal regime:
\bq
\beta^{(a)}_{\rm conf}(a_{*})=\beta^{({\ov a})}_{\rm conf}({\ov a}_{*}, a^{*}_f)=\beta^{({a_f})}_{\rm conf}({\ov a}_{*}, a^{*}_f)=0\,\,\ra\,\, \gamma_Q(a_{*})=\frac{3N_c-N_F}{N_F},\quad \gamma_{\Phi}(a_{*})= -2 \gamma_Q(a_{*})\,, \label{(1.5)}
\eq
\bbq
\gamma_q({\ov a}_{*}, a^{*}_f)=\frac{3\nd-N_F}{N_F},\quad \gamma_{M}({\ov a}_{*}, a^{*}_f)= - 2 \gamma_q({\ov a}_{*}, a^{*}_f)\,,
\eeq
in the direct and dual theories respectively.

\section{Quark and gluino condensates and multiplicities of vacua at $\mathbf{3N_c/2<N_F<2N_c}$}

\hspace{3mm}  To obtain the numerical values of the quark condensates (really, the mean vacuum values) $\langle{\ov Q}_j Q^i\rangle=\delta^i_j\langle ({\ov Q}Q)_i\rangle$, {\bf but only for this purpose}, the simplest way is to use the known {\bf exact form} of the non-perturbative contribution $\cw_{\rm non-pert}$ to the effective superpotential in the standard SQCD (i.e. without the fion fields $\Phi$). It seems clear that at sufficiently large values of $\mph\gg\la$ among the vacua of the $\Phi$-theory there will be $N_c$ vacua of the standard SQCD in which, definitely, all fions $\Phi$ are too heavy and dynamically irrelevant. Therefore, they all can be integrated out and this only results in additional 4-quark term in the superpotential, so that {\bf the exact} effective superpotential will look as
\bq
{\cw}_{\rm eff}=\Biggl [\cw_{\rm non-pert}=\,-\nd S=\,-\nd\Bigl (\frac{\det {\ov Q}Q}{\la^{\bo}}\Bigr )^{1/\nd}\Biggr ]+m_Q{\rm Tr}\,\qq -\frac{1}{2\mph}\Biggl [{\rm Tr} ({\qq})^2- \frac{1}{N_c}({\rm Tr}\,\qq)^2  \Biggr ], \,\,\,\label{(2.1)}
\eq
where the first non-perturbative term in \eqref{(2.1)} is well known in the standard ${\cal N}=1$ SQCD without fions.

Indeed, e.g. at $3N_c/2<N_F<2N_c$ and sufficiently large $\mph$, there are $N_c$ SQCD vacua in \eqref{(2.1)} with the unbroken $U(N_F)$ global flavor symmetry. In these, the last 4-quark term in \eqref{(2.1)} gives a small correction only and can be neglected and one obtains the well known results
\bq
\langle{\ov Q}_j Q^i\rangle_{SQCD}\approx\delta^i_j\frac{1}{m_Q}\Bigl (\lym^{(\rm SQCD)}\Bigr )^3=\delta^i_j\frac{1}{m_Q}\Bigl (\la^{\bo}m_Q^{N_F}\Bigr)_{,}^{1/N_c}\,\,\, \langle S\rangle_{SQCD}=\langle\frac{\lambda\lambda}{32\pi^2}\rangle_{SQCD}\approx \Bigl (\la^{\bo}m_Q^{N_F}\Bigr)^{1/N_c}.\,\,\, \label{(2.2)}
\eq
Now, using the holomorphic dependence of the superpotential \eqref{(2.1)} on the chiral superfields $({\ov Q}_j Q^i)$ and the chiral parameters $m_Q$ and $\mph$, the exact form \eqref{(2.1)} can be used to find the values of the quark condensates $\langle{\ov Q}_j Q^i\rangle$ and multiplicities of vacua in all other numerous vacua of the $\Phi$-theory and at all other values of $\mph\gg\la$. It is worth recalling only that, in general, as in the standard SQCD without additional fields $\Phi^i_j$, ${\cw}_{\rm eff}$ in \eqref{(2.1)} {\bf is not the superpotential of the genuine low energy Lagrangian describing lightest particles, it determines only the values of the vacuum mean values} $\langle{\ov Q}_j Q^i\rangle$ and $\langle S\rangle$. (The genuine low energy Lagrangians will be obtained below, both in the direct and dual theories).

Useful relations from \eqref{(2.1)} for vacua with broken flavor symmetry look as
\bq
\langle\Qo+\Qt-\frac{1}{N_c}{\rm Tr}\,(\qq)\rangle=m_Q\mph.\quad \langle S\rangle=\Biggl (\frac{\det\langle\qq\rangle}{\la^\bo}\Biggr )^{1/\nd}
=\frac{\langle\Qo\rangle\langle\Qt\rangle}{\mph}.\label{(2.3)}
\eq

It follows from \eqref{(2.1)},\eqref{(2.3)} that there is a large number of various different vacua in this theory. But as for a realization of the global flavor symmetry $U(N_F)$, there are only two types of vacua: those with unbroken $U(N_F)$ and those with the spontaneous breaking $U(N_F)\ra U(\no)\times U(\nt),\, \no+\nt=N_F,\, \no\leq N_F/2$\,.

As an example, we consider below only the br2-vacua (br=breaking) with $\langle\Qt\rangle\gg\langle\Qo\rangle$ and $\nt>N_c,\, \no<\nd$, and with the multiplicity
$N_{\rm br2}=(\nd-\no)C^{\no}_{N_F},\,\, C^{\no}_{N_F}=N_F!/(n_1!\, n_2!)$.

\section{Fions $\mathbf{\Phi^i_j}$ in the direct theory\,: one or three generations}

At all those scales $\mu<\la$ until the field $\Phi$ remains too heavy and non-dynamical (while the light quarks and gluons are still effectively massless and dynamical), i.e. until the perturbative running mass $\mu_{\Phi}^{\rm pert}(\mu)>\mu$, the field $\Phi$ decouples and can be integrated out, and the Lagrangian in the conformal regime takes the form at the scale $\mu\ll\la$ (${\ov Q_{R}}, Q_{R}$ are renormalized fields)
\bbq
\hspace*{-4mm} K=z_Q(\la,\mu){\rm Tr}\Bigl (Q^\dagger Q+Q\ra {\ov Q}\Bigr )={\rm Tr}\Biggl (Q_{R}^\dagger Q_{R}+(Q_{R}\ra {\ov Q_{R}})\Biggr ),\,\, z_Q(\la,\mu\ll\la)=\Bigl (\frac{\mu}{\la}\Bigr )^{\gamma_Q=\frac{3N_c-N_F}{N_F}\,>\,0}\ll 1\,,
\eeq
\vspace*{-4mm}
\bq
\cw_Q=\frac{m_Q}{z_Q(\la,\mu)}{\rm Tr}\Bigl ({\ov Q_{R}} Q_{R}\Bigr )-\frac{1}{2\mph z^2_Q(\la,\mu)}\Biggl ({\rm Tr}\,({\ov Q}_{R}Q_{R})^2-\frac{1}{N_c}\Bigl({\rm Tr}\,{\ov Q}_{R} Q_{R} \Bigr)^2 \Biggr ).\,\,\label{(3.1)}
\eq

Because the quark renormalization factor $z_Q(\la,\mu)$ decreases at smaller scale $\mu$, it is seen  from \eqref{(3.1)} that the role of the 4-quark term $\sim ({\ov Q}_R Q_R)^2$ increases with lowering energy. Hence, while it is irrelevant at the scale $\mu\sim\la$ because $\mph\gg \la$, the question is whether it becomes dynamically relevant at some lower scale $\mu=\mu_o$. For this, we estimate the scale $\mu_o$ where this term becomes relevant in the conformal regime of the (effectively) massless theory of quarks and gluons, with $z_Q(\la,\mu\ll\la)\sim (\mu/\la)^{\gamma_Q}$\,:
\bq
\frac{\mu_o}{\mph}\frac{1}{z^2_Q(\la,\mu_o)}\sim\frac{\mu_o}{\mph}\Bigl (\frac{\la}{\mu_o}\Bigr )^{2\gamma_Q}\sim 1\quad\ra \quad \mu_o\sim \la\Bigl (\frac{\la}{\mph}\Bigr)^{\frac{1}{(2\gamma_Q-1)}}\sim\la\Bigl (\frac{\la}{\mph}\Bigr )^{\frac{N_F}{3(2N_c-N_F)}\,>\,0}\ll\la\,. \label{(3.2)}
\eq

We recall that even at those scales $\mu$ when the running mass of fions $\mu_{\Phi}(\mu)=\mph/z_{\Phi}(\la,\mu)\gg \mu$ and so they are too heavy and dynamically irrelevant, {\bf the quarks and gluons remain effectively massless and active}. Therefore, due to the Yukawa interactions of fions with quarks, the loops  of still active light quarks (and gluons interacting with quarks) {\bf still induce the power-like running renormalization factor $z_{\Phi}(\la,\mu\ll
\la)\sim (\mu/\la)^{\gamma_{\Phi}<0}\gg 1$ of fions at all those scales until quarks are effectively massless}, i.e. $\mu>m^{\rm pole}_Q$ (see below).

It seems clear that the physical reason why the $4$-quark terms in the superpotential \eqref{(3.1)} become relevant at scales $\mu<\mu_o$ is that {\bf the fion field $\Phi$ which was too heavy and so dynamically irrelevant at $\mu>\mu_o,\, \mph(\mu>\mu_o)>\mu$\,, becomes effectively massless at $\mu<\mu_o,\, \mph(\mu<\mu_o)<\mu$\,, and begins to participate in the renormgroup evolution, i.e. it becomes relevant}. In other words, the 4-quark term in \eqref{(3.1)} `remembers' about fions and signals about the scale below which the fions become effectively massless, $\mu_o=\mu^{\rm pole}_2(\Phi)$. This allows us to find the value of $z_{\Phi}(\la,\mu>\mu_o)$:
\bq
\frac{\mph}{z_{\Phi}(\la,\mu_o)}\sim\mu_o\quad\ra\quad z_{\Phi}(\la,\mu_o<\mu<\la)\sim\Bigl (\frac{\la}{\mu}\Bigr )^{2\gamma_Q}\gg 1\quad\ra\quad \gamma_{\Phi}=-2\gamma_Q<0\,. \label{(3.3)}
\eq

Because the propagator of the renormalized fion field looks as $1/(p^2-\,\mu^2_{\Phi}(p^2))$ and $|\mu^2_{\Phi}(p^2)|\lessgtr |p^2|$ at $p^2\lessgtr\mu^2_o$, where $\mu_o\ll\la$ \eqref{(3.2)}, it is clear that there is a pole in the fion propagator at $p^2=\mu_{2}^{\rm pole}(\Phi)=(\mu^2_o-i\mu_o\Gamma_{\Phi})$, i.e. {\bf there is a second generation of all $N_F^2$ fields $\Phi^i_j$} (the first one is at $\mu^{\rm pole}_1(\Phi)\gg\la$).

It can be shown that {\bf the conformal regime remains the same} even at scales $m^{\rm pole}_Q<\mu<\mu_o$ where  fion fields became relevant, and the quark and fion anomalous dimensions $\gamma_Q$ and $\gamma_{\Phi}$ remain the same. I.e., the perturbative running mass $\mph(\mu)=\mph/z_{\Phi}(\la,\mu\ll\la)\ll\la$ of fions continues to decrease quickly with diminishing $\mu$ at all scales $m^{\rm pole}_Q<\mu<\la$ until quarks remain effectively massless, and becomes frozen only at scales below the quark physical mass $m^{\rm pole}_Q$, when the heavy quarks decouple.\\

However, if $m^{\rm pole}_Q >\mu_o$\,, there is no pole in the fion propagator at scales $\mu<\la$. The reason is that quarks decouple as heavy at $\mu<m^{\rm pole}_Q$. And because $m^{\rm pole}_Q >\mu_o$, all fions $\Phi^i_j$ remain too heavy and irrelevant at this scale. Then, at $\mu<m^{\rm pole}_Q$, the running fion mass remains frozen at the large value $\mph(\mu=m^{\rm pole}_Q>\mu_o)>m^{\rm pole}_Q$. The fions remain then dynamically irrelevant and unobservable as
resonances in this case at all scales $\mu<\la$.

But when $m^{\rm pole}_Q\ll\mu_o$, there will be not only the second generation of fions at $\mu=\mu^{\rm pole}_2(\Phi)$, but also {\bf a third generation} at $\mu=\mu^{\rm pole}_3(\Phi)\ll\mu^{\rm pole}_2(\Phi)$. Indeed, after the heavy quarks decouple at the scale $m^{\rm pole}_Q\ll\mu_o$ and the renormalization factor $z_{\Phi}(\la,m^{\rm pole}_Q)$ of fions becomes frozen {\bf in the region of scales where the fions already became relevant}, the frozen value $\mph(\mu<m^{\rm pole}_Q)=\mph/z_{\Phi}(\la,\mu=m^{\rm pole}_Q)$ of the fion mass is now: $\mph(\mu= m^{\rm pole}_Q)\ll m^{\rm pole}_Q$. Therefore, {\bf there is one more pole in the fion propagator} at $\mu=\mu^{\rm pole}_3(\Phi)=\mph(\mu = m^{\rm pole}_Q)\ll m^{\rm pole}_Q$.\\

On the whole, in a few words for the direct theory.\\
{\bf a)} The fions remain dynamically irrelevant and there are no poles in the fion propagator at scales $\mu<\la$ if $m^{\rm pole}_Q>\mu_o$.\\
{\bf b)} If $m^{\rm pole}_Q<\mu_o\sim\la\Bigl (\la/\mph\Bigr )^{\frac{N_F}{3(2N_c-N_F)}}\ll\la$, there are two poles in the fion propagator at scales $\mu\ll\la$\,:\, $\mu^{\rm pole}_2(\Phi)\sim\mu_o$ and $\mu^{\rm pole}_3(\Phi)=\mph/z_{\Phi}(\la,m^{\rm pole}_Q)\ll\mu^{\rm pole}_2(\Phi)$. In other words, {\bf the fions appear in three generations} in this case (we recall that there is always the largest pole mass of fions $\mu^{\rm pole}_1(\Phi)\sim\mph\gg\la$). Hence, the fions are effectively massless and dynamically relevant in the range of scales $\mu^{\rm pole}_3(\Phi)<\mu<\mu^{\rm pole}_2(\Phi)$.

Moreover, once the fions become relevant with respect to internal interactions, they begin to contribute simultaneously to the external anomalies ( the 't Hooft triangles in the external background fields).
\vspace*{-3mm}

\section{ Mass spectra in $\rm br2$ vacua. Direct theory}
\vspace*{-0.1cm}
\hspace*{0.5cm} $\bd/N_F=O(1),\,\,0<(\bd-2\no)/N_F=O(1), \,\,\, \la\ll\mph\ll\mo=\la (\la/m_Q)^{(2N_c-N_F)/N_c}$\\

The general scheme for calculations of mass spectra both in the direct and dual theories looks as follows.

1) From the exact ${\cw}_{\rm eff}$ in \eqref{(2.1)} the values of the quark and gluino condensates at $\mu=\la$, $\langle ({\ov Q} Q)_i\rangle$ and $\langle S\rangle$, as well as multiplicities of vacua, can be found in each vacuum.

2) From this and from the knowledge of all anomalous dimensions in the conformal regime, all renormalization factors $z_i(\la,\mu\ll\la)$ for all fields in the Kahler terms are also known. Then the potentially possible values of pole masses of quarks, $m^{\rm pole}_Q=\langle m_Q^{\rm tot}\rangle/z_Q(\la,m^{\rm pole}_Q)$, or possible gluon pole masses $(\mu^{\rm pole}_{gl})^2\sim z_Q(\la,\mu_{gl})\langle{\ov Q}\rangle\langle Q\rangle$ for higgsed quarks can be found (and, using the Konishi anomalies \cite{Konishi} and matching $\langle M^i_j\rangle=\langle{\ov Q}_j Q^i\rangle,\, \langle S\rangle= - \langle{\ov S}\rangle$ similarly in the dual theory).

3) The hierarchies between them determine then the realized phase states and real mass spectra in each vacuum at given values of Lagrangian parameters.
E.g., if (see below) for dual quarks with $U(\no)$ flavors ${\ov\mu}^{\rm pole}_{\rm gl,1}\gg \mu^{\rm pole}_{q,1}$, then these quarks are higgsed, i.e.
$\langle\qo\rangle=\sum_{a=1}^{\nd}\langle {\ov q}^1_a q^a_1\rangle=\langle {\ov q}^1_1\rangle\langle q^1_1\rangle\sim m_Q\la$, and the dual color symmetry is broken: $SU(\nd)\ra SU(\nd-\no)$. While if for quarks ${\ov Q}^a_1, Q^1_a$ in the direct theory $m^{\rm pole}_{Q,1}\gg \mu^{\rm pole}_{gl,1}$, then these quarks decouple as heavy at $\mu<m^{\rm pole}_{Q,i}$ and are not higgsed but confined. The confinement originates from the unbroken color $SU(N_c)\,\, {\cal N}=1$ supersymmetric YM (SYM) with its only dimensional parameter $\langle\lym\rangle=\langle S\rangle^{1/3}$, so that the string tension is $\sigma^{1/2}\sim\langle\lym\rangle$.

\hspace*{4mm} From ${\cw}_{\rm eff}$ in \eqref{(2.1)},\eqref{(2.3)} the condensates of quarks in the direct theory look as ($m_1=m_Q N_c/(N_c-\nt)$\,):
\bq
\hspace*{-6mm}\langle\Qt\rangle= m_1\mph-\frac{N_c-\no}{N_c-\nt}\langle\Qo\rangle,\,\, \langle\Qo\rangle\approx\la^2\Bigl(\frac{\mph}{\la}\Bigr )^{\frac{n_2}{n_2-N_c}}\Bigl (\frac{m_1}{\la}\Bigr )^{\frac{N_c-n_1}{n_2-N_c}},\,\, \frac{\langle\Qo\rangle}{\langle\Qt\rangle}\sim\biggl (\frac{\mph}{\mo}\biggr )^{\frac{N_c}{\nt-N_c}}\ll 1,\,\,\,\,\label{(4.1)}
\eq
in br2 - vacua with the spontaneous breaking $U(N_F)\ra U(\no)\times U(\nt)$, \,$n_2>N_c\,, 1\leq n_1<\nd$\,. The largest among the masses smaller than $\la$ are {\bf masses of $N_F^2$ second generation fions}, see \eqref{(3.2)},
\bq
\mu^{\rm pole}_2(\Phi_i^j)=\mu_o\sim\la\Bigl (\frac{\la}{\mph}\Bigr )^{\frac{N_F}{3(2N_c-N_F)}}\ll\la\,, \quad i,j=1...N_F\,,\label{(4.2)}
\eq
and {\bf all $N_F^2$ fions become dynamically relevant at scales $\mu<\mu_o$} (the cases when there are additional  non-perturbative contributions to the masses of fions have to be considered separately, see below).

Some other possible characteristic masses look in this vacuum as
\footnote{\,
Here and below, $m^{\rm pole}_{Q,1},\,m^{\rm pole}_{Q,2}$ in the direct theory and ${\ov\mu}^{\rm pole}_{\rm gl,1},\, \mu^{\rm pole}_{\dq,2}$ in the dual one are the pure perturbative pole masses of quarks or gluons, i.e. ignoring confinement with the string tension $\sigma^{1/2}\sim\langle\lym\rangle$.
}
\bq
\langle m^{\rm tot}_{Q,1}\rangle=\frac{\langle\Qt\rangle}{\mph}\approx m_1\,,\quad  m^{\rm pole}_{Q,2}\ll m^{\rm pole}_{Q,1}=\frac{\langle m^{\rm tot}_{Q,1}\rangle}{z_Q(\la,m^{\rm pole}_{Q,1})}\sim\la\Bigl(\frac{m_1}{\la}\Bigr )^{N_F/3N_c}\ll\mu^{\rm pole}_2(\Phi_i^j)\,, \label{(4.3)}
\eq
\vspace*{-4mm}
\bq
\hspace*{-4mm}\mgt^2\sim z_Q(\la,\mgt)\langle\Qt\rangle\gg\mgo^2,\,\, z_Q(\la,\mgt)\sim\Bigl (\frac{\mgt}{\la}\Bigr )^{\frac{3N_c-N_F}{N_F}}\ll 1\,\ra\,  \mgt\sim\langle\lym\rangle\ll m^{\rm pole}_{Q,1},\,\label{(4.4)}
\eq
where $m^{\rm pole}_{Q,1}$ and $m^{\rm pole}_{Q,2}$ are the pole masses of quarks ${\ov Q}_1, Q^1$ and ${\ov Q}_2, Q^2$ and $\mgo,\, \mgt$ are the gluon masses due to possible higgsing of these quarks. Hence, the largest mass is $m^{\rm pole}_{Q,1}$. {\bf The overall phase is: all heavy quarks}, i.e. not higgsed but confined, $\langle {\ov Q}_1\rangle=\langle Q^1\rangle=\langle {\ov Q}_2\rangle=\langle Q^2\rangle=0$.\\

After the heaviest quarks $Q^1,\,{\ov Q}_1$ decoupled at $\mu<m^{\rm pole}_{Q,1}$,  the lower energy theory has $N_c$ colors and $N_F^\prime =n_2>N_c$ flavors of still active lighter quarks ${\ov Q}_2, Q^2$. In the range of scales $m^{\rm pole}_{Q,2}<\mu<m^{\rm pole}_{Q,1}$ it will remain in the conformal regime at $2n_1<\,\bd,\,\,\bd=(3\nd-N_F)>0$, while it will be not in the conformal but in the strong coupling regime at $2n_1>\bd$, with the gauge coupling $a(\mu\ll m^{\rm pole}_{Q,1})=(m^{\rm pole}_{Q,1}/\mu)^{\nu\,>\,0}\gg 1$. We do not consider the strong coupling regime here and for this reason we consider $2n_1<\bd$ only.\\

It follows from the exact ${\cw}_{\rm eff}$ in \eqref{(2.1)} that the flavor symmetry is broken spontaneously in these br2 vacua as $U(N_F)\ra U(\no)\times U(\nt)$. It follows then from this that quarks ${\ov Q}_2, Q^2$ are not higgsed. If they were higgsed, then $U(\nt)$ would be further broken spontaneously due to the rank restriction because $\nt>N_c$, this would contradict the exact \eqref{(2.1)}. Therefore the quarks ${\ov Q}_2, Q^2$ in this case are not higgsed but confined.

In the lower energy theory at $\mu<m^{\rm pole}_{Q,1}$ the pole mass of quarks ${\ov Q}_2, Q^2$ looks as
\bq
m^{\rm pole}_{Q,2}=\frac{m^{\rm pole}_{Q,1}}{z^{\,\prime}_Q(m^{\rm pole}_{Q,1},m^{\rm pole}_{Q,2})}\Biggl (\,\frac{\langle\Qo\rangle}{\langle\Qt\rangle}\Biggr )\sim (\rm several)\lym,  \quad z^{\,\prime}_Q(m^{\rm pole}_{Q,1},m^{\rm pole}_{Q,2})\sim\Bigl (\frac{m^{\rm pole}_{Q,2}}{m^{\rm pole}_{Q,1}}\Bigr )^{\frac{3N_c-n_2}{n_2}}\ll 1\,. \label{(4.5)}
\eq

Hence, after integrating out  as heavy the quarks ${\ov Q}_1, Q^1$ at $\mu<m^{\rm pole}_{Q,1}$ and then  quarks ${\ov Q}_2, Q^2$ and $SU(N_c)$ gluons at $\mu<\langle\lym\rangle$ (these last through the Veneziano - Yankielowicz procedure \cite{VY}), the Lagrangian of fions looks as, see \eqref{(4.5)},
\bq
K=z_{\Phi}(\la,m^{\rm pole}_{Q,1})\,{\rm Tr}\,\Bigl [\,(\Phi_1^1)^\dagger \Phi_1^1+(\Phi_1^2)^\dagger \Phi_1^2+(\Phi_2^1)^\dagger \Phi_2^1+z^{\,\prime}_{\Phi}(m^{\rm pole}_{Q,1},m^{\rm pole}_{Q,2})(\Phi_2^2)^\dagger \Phi_2^2\,\Bigr ]\,, \label{(4.6)}
\eq
\vspace*{-3mm}
\bq
z_{\Phi}(\la,m^{\rm pole}_{Q,1})\sim\Bigl (\frac{\la}{m^{\rm pole}_{Q,1}}\Bigr )^{\frac{2(3N_c-N_F)}
{N_F}}\gg 1,\quad  \cw=N_c S+\cw_{\Phi},\quad \cw_{\Phi}=\frac{\mph}{2}\Bigl ({\rm Tr}\,(\Phi^2)-
\frac{1}{\nd}({\rm Tr}\,\Phi)^2 \,\Bigr )\,,  \label{(4.7)}
\eq
\bbq
\langle m^{\rm tot}_{Q,1}\rangle=\frac{\langle\Qt\rangle}{\mph}\,,\quad \langle m^{\rm tot}_{Q,2}\rangle=\frac{\langle\Qo\rangle}{\mph}\,,\quad  m^{\rm tot}_Q=(m_Q-\Phi)\,,\quad S=\Bigl (\la^{\bo}\det m^{\rm tot}_Q\Bigr )^{1/N_c}\,,
\eeq
\bbq
\langle\lym\rangle^3=\langle S\rangle=\Bigl (\la^{\bo}\det \langle m^{\rm tot}_Q\rangle\Bigr )^{1/N_c}\approx\la^3\Bigl (\frac{\mph}{\la}\Bigr )^{\frac{n_2}{n_2-N_c}}\Bigl (\frac{m_1}{\la}\Bigr )^{\frac{\nt-n_1}{n_2-N_c}},\quad m_1=\frac{N_c m_Q}{N_c-\nt}.
\eeq

From \eqref{(4.6)},\eqref{(4.7)}, the main contribution to the mass of $\no^2$ {\bf third generation fions} $\Phi_1^1$ gives the term $\sim\mph (\Phi_1^1)^2$,
\vspace*{-5mm}
\bq
\mu^{\rm pole}_3(\Phi_1^1)=\frac{\mph}{z_{\Phi}(\la,m^{\rm pole}_{Q,1})}\sim \mph\Bigl (\frac{m_Q}{\la}\Bigr )^{\frac{2(3N_c-N_F)}{3N_c}}
\sim \Bigl (\frac{\mph}{\mo}\Bigr )^{\frac{(\bd-2\no)}{3(\nt-N_c)}\,>\,0}\langle\lym\rangle\ll\langle\lym\rangle\,.\label{(4.8)}
\eq

As for {\bf $\nt^2$ third generation fions} $\Phi_2^2$, the main contribution to their masses comes from the non-perturbative term $\sim S$ in the superpotential \eqref{(4.7)}
\bq
\mu^{\rm pole}_3(\Phi_2^2)=\frac{\langle S\rangle}{\langle m^{\rm tot}_{Q,2}\rangle^2}\frac{1}
{z_{\Phi}(\la,m^{\rm pole}_{Q,1}) z^{\,\prime}_{\Phi}(m^{\rm pole}_{Q,1},m^{\rm pole}_{Q,2})}\sim m^{\rm pole}_{Q,2}\sim\langle\lym\rangle. \label{(4.9)}
\eq

{\bf $2\no\nt$ third generation hybrid fions $\Phi_1^2, \Phi_2^1$ are massless}: $\mu^{\rm pole}_3(\Phi_1^2)=\mu^{\rm pole}_3(\Phi_2^1)=0$, they are Nambu-Goldstone particles of the spontaneously broken global flavor symmetry: $U(N_F)\ra U(\no)\times U(\nt)$.

\section{ Mass spectra in $\rm br2$ vacua. Dual theory}
\vspace*{-0.1cm}
\hspace*{0.5cm} $\bd/N_F=O(1),\,\,0<(\bd-2\no)/N_F=O(1), \,\,\, \la\ll\mph\ll\mo=\la (\la/m_Q)^{(2N_c-N_F)/N_c}$\\

In these vacua with $n_2>N_c\,, 1\leq n_1<\nd$, using the Konishi anomalies \cite{Konishi} and matching $\langle M^i_j\rangle=\langle{\ov Q}_j Q^i\rangle,\, \langle S\rangle= - \langle{\ov S}\rangle$, see also \eqref{(4.7)}, the condensates of mions and dual quarks look at $\mu=\la$ as:
\bq
\langle M_2\rangle=\langle\Qt\rangle\approx m_1\mph,\,\,\langle M_1\rangle=\langle\Qo\rangle\approx\la^2\Bigl(\frac{\mph}{\la}\Bigr )^{\frac{n_2}{n_2-N_c}}\Bigl (\frac{m_1}{\la}\Bigr )^{\frac{N_c-n_1}{n_2-N_c}},\,\,\frac{\langle M_1\rangle}{\langle M_2\rangle}\sim\Bigl (\frac{\mph}{\mo}\Bigr )^{\frac{N_c}{n_2-N_c}}\ll 1,\,\,\label{(5.1)}
\eq
\bbq
\langle\lym\rangle^3=-\langle{\ov S}\rangle=\frac{\langle\Qo\rangle\langle\Qt\rangle}{\mph}=\frac{\langle M_1\rangle\langle M_2\rangle}{\mph},\,\,\langle N_1\rangle=\langle\qo\rangle=\frac{\la\langle S\rangle}{\langle M_1\rangle}=\frac{\la\langle M_2\rangle}{\mph}\approx m_1\la\gg\langle \qt\rangle.
\eeq

From these and \eqref{(1.5)}, the heaviest are $N_F^2$ mions $M^i_j$ with the pole masses
\bq
\mu^{\rm pole}(M^i_j)=\frac{\la^2/\mph}{z_M(\la,\mu^{\rm pole}(M))}\sim \la\Bigl (\frac{\la}{\mph}\Bigr )^{\frac{N_F}{3(2N_c-N_F)}}\sim \mu^{\rm pole}_2(\Phi_i^j)\gg{\ov\mu}^{\rm pole}_{\rm gl,1}\,, \label{(5.2)}
\eq
\bbq
z_M(\la,\mu^{\rm pole}(M))\sim\Bigl(\frac{\mu^{\rm pole}(M)}{\la}\Bigr )^{\gamma_M= -2\gamma_q= -2\frac{3\nd-N_F}{N_F}}\gg 1\,,
\eeq

\hspace*{-8mm} while some other possible characteristic masses look as
\bq
\Bigl ({\ov\mu}^{\rm pole}_{\rm gl,1}\Bigr )^2\sim z_q(\la,{\ov\mu}^{\rm pole}_{\rm gl,1})\langle {\ov q}^1_1\rangle\langle q^1_1\rangle,
\quad {\ov\mu}^{\rm pole}_{\rm gl,1}\sim \la\Bigl(\frac{m_Q}{\la}\Bigr )^{\frac{N_F}{3N_c}}\sim m^{\rm pole}_{Q,1}\gg{\ov\mu}^{\rm pole}_{\rm gl,2}, \quad
{\ov\mu}^{\rm pole}_{\rm gl,1}\gg \mu^{\rm pole}_{q,2}\gg\mu^{\rm pole}_{q,1},\,\,\label{(5.3)}
\eq
where ${\ov\mu}^{\rm pole}_{\rm gl,1,2}$ are the gluon masses due to possible higgsing of these quarks. Hence, the largest mass is ${\ov\mu}_{\rm gl,1}$ and the overall phase is ${\bf Higgs_1-Hq_2}$ (i.e. higgsed quarks $q_1$ and confined quarks $\dq_2$ with non-higgsed colors). The quarks ${\odq}^2,\,\dq_2$ with the $U(\nt>N_c)$ flavor symmetry are not higgsed due to the same rank restriction as the quarks ${\ov Q}_2,\, Q^2$ of the direct theory.

After integrating out all massive gluons and their scalar superpartners, the dual Lagrangian at $\mu=\muo$ looks as
\bq
K= z_M(\la,\muo){\rm Tr}\,\frac{M^\dagger M}{\la^2}+ z_q(\la,\muo){\rm Tr}\,\Bigl [\,2\sqrt{(N_1^1)^\dagger N_1^1}+K_{\rm hybr}+\Bigl ( (\dq_2)^{\dagger}\dq_2 +({\dq}_2\ra {\odq}^{\,2} )\Bigr )\,\Bigr ]\,,\,\, \label{(5.4)}
\eq
\bbq
K_{\rm hybr}=\Biggl ( (N_1^2)^{\dagger}\frac{1}{\sqrt{N_1^1 (N_1^1)^{\dagger}}} N_1^2+(N_1^2\ra N_2^1)
\Biggr ),\,\, z_q(\la,\muo)\sim\Bigl (\frac{\muo}{\la}\Bigr )_{,}^{\frac{\bd}{N_F}}\ll 1,\,\, z_M=z^{-2}_q, \,\, \bd=3\nd-N_F,
\eeq
\bq
{\ov{\cal W}} =\Bigl [-\frac{2\pi}{{\ov\alpha}(\mu)}{\ov{\textsf S}}\Bigr ]-{\rm Tr}\,\Bigl ({\odq}^2\frac{ M_2^2}{\la}{\dq}_2\Bigr )+ \cw_{MN}+\cw_{M}, \,\, \cw_M=m_Q{\rm Tr}\,M -\frac{1}{2\mph}\Biggl [{\rm Tr}\, (M^2)- \frac{1}{N_c}({\rm Tr}\, M)^2 \Biggr ],\,\,\label{(5.5)}
\eq
\bbq
\cw_{MN}=\frac{-1}{\la}\,{\rm Tr}\,\Bigl (M_1^1 N_1^1+M_2^1 N_1^2+ M_1^2 N_2^1+M_2^2 N_2^1\frac{1}{N_1^1} N_1^2\Bigr ),\,\, N^1_2=(\langle{\ov q}^1\rangle q_2),\,\, N^2_1=({\ov q}^2\langle q_1\rangle),
\eeq
where $\no^2$ nions (dual pions) $N_1^1$ originate from higgsing of ${\ov q}^1, q_1$ dual quarks, while the hybrid nions $N_1^2$ and $N_2^1$ are, in essence, the dual quarks ${\ov q}^2$ and $q_2$ with higgsed colors. ${\odq}^2, \dq_2$ are still active quarks ${\ov q}^2,\,q_2$ with non-higgsed colors. $\ov{\textsf S}$ is the field strength squared of remained light dual $SU(\nd-\no)$ gluons.

The lower energy theory at $\mu<\muo$ has $(\nd-n_1)$ colors and $n_2>N_c$ flavors, $0<\bd^{\,\prime}=(\bd-2n_1)\,<\,\bd$. We consider here only the case $\bd^{\,\prime}>0$ when it remains in the conformal window. The fields $N_1^1, N_1^2, N_2^1$ and $M_1^1, M_1^2, M_2^1$ are frozen and do not evolve at $\mu<\muo$, while the value of the pole mass $\qtp$ in this lower energy theory is
\bq
\mu^{\rm pole}_{\dq,2}=\frac{\langle M_2\rangle}{\la}\frac{1}{z_q(\la,\muo) z^{\,\prime}_q(\muo,\mu^{\rm pole}_{\dq,2})}\sim (\rm several)\langle
\lym\rangle\,, \quad z^{\,\prime}_q(\muo,\mu^{\rm pole}_{\dq,2})\sim\Bigl (\frac{\mu^{\rm pole}_{\dq,2}}{\muo}\Bigr )^{\bd^{\,\prime}/n_2}\ll 1\,. \label{(5.6)}
\eq

Finally, after integrating out remained non-higgsed (but confined) quarks ${\odq}^2, \dq_2$ (confinement originates in this case from the $SU(\nd-\no)\,\, {\cal N}=1$ SYM sector with its scale factor $\langle\lym\rangle$) as heavy ones and then ${\cal N}=1$ $SU(\nd-\no)$ SYM gluons at $\mu<\langle\lym\rangle$ (these last through the Veneziano - Yankielowicz procedure \cite{VY}), the lowest energy Lagrangian of mions and nions looks as, see \eqref{(5.4)},
\bq
K=z_M(\la,\muo){\rm Tr}\,K_M+z_q(\la,\muo) {\rm Tr}\,\Bigl [\,2\sqrt{(N_1^1)^\dagger N_1^1}+K_{\rm hybr}\,\Bigr ],\,\,\,\,\label{(5.7)}
\eq
\bbq
K_M=\frac{1}{\la^2}\Bigl ( (M_1^1)^\dagger M_1^1+(M_1^2)^\dagger M_1^2+(M_2^1)^\dagger M_2^1+
z^{\,\prime}_M (\muo,\mu^{\rm pole}_{\dq,2}) (M_2^2)^\dagger M_2^2\Bigr ),\,\, z^{\,\prime}_M (\muo,\mu^{\rm pole}_{\dq,2})\sim\Bigl (\frac{\muo}{\mu^{\rm pole}_{\dq,2}}\Bigr )^{\frac{2\,\bd^{\,\prime}}{n_2}}\gg 1,
\eeq
\bbq
\cw=-\nd^{\,\prime}\,{\cal S}+\cw_{MN}+\cw_M,\,\, {\cal S}=\langle\lym\rangle^3\Biggl (\det\frac{\langle N_1\rangle}{N_1^1}\det\frac{M_2^2}
{\langle M_2\rangle}\Biggr )^{1/\nd^{\,\prime}},\,\, \nd^{\,\prime}=(\nd-\no),\,\, \langle\lym\rangle^3\approx m_1\langle M_1\rangle.
\eeq

From \eqref{(5.7)}, the "masses" of mions at the low scale look as
\bq
\mu_{low}(M_1^1)\sim\mu_{low}(M_1^2)\sim\mu_{low}(M_2^1)\sim\frac{\la^2}{z_M(\la,\muo)\mph}\sim\Bigl (\frac{\mo}{\mph}\Bigr )\muo\gg\muo\,,\quad \frac{\mu_{low}(M_1^1)}{\mu^{\rm pole}(M)}\ll 1\,,\label{(5.8)}
\eq
\vspace*{-4mm}
\bq
\mu_{low}(M_2^2)\sim\frac{\la^2}{z_M(\la,\muo)z^\prime_M (\muo,\mu^{\rm pole}_{\dq,2})\mph}\sim \Bigl (\frac{\mo}{\mph}\Bigr )^{\frac{3N_c-n_2}{3(n_2-N_c)}}\,\muo\gg \muo\,,\quad \frac{\mu_{low}(M_2^2)}{\mu_{low}(M_1^1)}\ll 1\,,\label{(5.9)}
\eq
while the pole masses of nions $N_1^1$ are
\bq
\mu^{\rm pole}(N_1^1)\sim \mph\Bigl (\frac{m_Q}{\la}\Bigr )^{\frac{2(3N_c-N_F)}{3N_c}}\sim\mu_{3}^{\rm pole}(\Phi^1_1)
\sim \Bigl (\frac{\mph}{\mo}\Bigr )^{\frac{(\bd-2\no)}{3(\nt-N_c)}\,>\,0}\langle\lym\rangle\ll\langle\lym\rangle\,.\label{(5.10)}
\eq

$2\no\nt$ hybrid nions $N_1^2, N_2^1$ {\bf are massless}: $\mu^{\rm pole}(N_1^2)=\mu^{\rm pole}(N_2^1)=0$, they are Nambu-Goldstone particles of the spontaneously broken global flavor symmetry: $U(N_F)\ra U(\no)\times U(\nt)$.

The large mion "masses" \eqref{(5.8)},\eqref{(5.9)} are not their pole masses but simply the frozen low energy values of their running masses. The reason is that all $N_F^2$ mion fields $M^i_j$ are light and dynamically relevant  only at scales $\mu^{\rm pole}(M)<\mu<\la$, see \eqref{(5.2)}. They become too heavy,  dynamically irrelevant and decouple at scales $\mu<\mu^{\rm pole}(M)$.  Nevertheless, their renormalization factors continue to grow with diminished energy due to couplings with lighter dual quarks. They become frozen for $M_1^1, M_1^2, M_2^1$ only at $\mu<\muo$ after the quarks ${\ov q}^1, q_1$ are higgsed, and at $\mu<\mu^{\rm pole}_{\dq,2}$ for $M_2^2$ after the quarks $\odq^2, \dq_2$ decouple as heavy. The only pole masses of all $N_F^2$ mions $M^i_j$ are $\mu^{\rm pole}(M)\sim \la\Bigl (\la/\mph\Bigr )^{N_F/3(2N_c-N_F)}$ in \eqref{(5.2)}.

\section{\quad Conclusions}

\hspace*{4mm} {\bf A). The qualitatively new phenomenon} was found in the direct theory due to the strong power-like renormgroup evolution in the conformal regime. - {\bf The seemingly heavy and dynamically irrelevant $N_F^2$ fion fields $\Phi_i^j$ `return back' and there appear two additional generations of light $\Phi$-particles with small masses $\mu_{3}^{\rm pole}(\Phi)\ll\mu_{2}^{\rm pole}(\Phi)\ll\la$}.
Moreover, the third generation fields $\Phi^1_2$ and $\Phi^2_1$ are massless, they are Nambu-Goldstone particles of the spontaneously broken global flavor symmetry $U(N_F)\ra U(\no)\times U(\nt)$.\\

{\bf B).} Let us compare now the mass spectra (for particle masses $M_k<\la$) in the direct theory and in Seiberg's dual one at $3N_c/2<N_F<2N_c$ and $\la\ll\mph\ll\mo=\la(\la/m_Q)^{(2N_c-N_F)/N_c}$.\\

\hspace*{2cm}{\bf Part I\,:\,\,Mass spectra at  $0<\bd/N_F=O(1)\,,\,0<(\bd- 2\no)/N_F=O(1)$}\\

1) The largest masses $\mu_2^{\rm pole}(\Phi^i_j)\sim\mu_o\sim\la(\la/\mph)^{N_F/3(2N_c-N_F)}$ in the direct theory have $N_F^2$ second generation scalar fion superfields $\Phi^i_j$ \eqref{(4.2)}, and $N_F^2$ scalar mion superfields $M^i_j$ \eqref{(5.2)} with parametrically the same pole masses in the dual one (here and below: up to possible constant factors independent of $m_Q$ and $\mph$ which are hard to control). Therefore, these two sets look undistinguishable (with our accuracy).

It is also worth noting that when all $N_F^2$ fion fields $\Phi^i_j$ become relevant at $\mu<\mu_o$ in the direct theory, then all $N_F^2$ mion fields $M^i_j$ become irrelevant in the dual one (and vice versa at $\mu>\mu_o$).\\

2) The next scale is $m_{Q,1}^{\rm pole}\sim \muo^{\rm pole}\sim\la(m_Q/\la)^{N_F/3N_c}\ll\mu_2^{\rm pole}(\Phi^i_j)$, \eqref{(4.3)},\eqref{(5.3)}. Because all quarks with $\no$ and $\nt$ flavors are confined in the direct theory and $m_{Q,1}^{\rm pole}\gg m_{Q,2}^{\rm pole}$, there are e.g.\,: a) {\bf many
flavored quarkonia with different spins, with this scale of masses, made either from quarks ${\ov Q}_1,\,Q^1$ with $\no$ flavors or e.g. from ${\ov Q}_1$ and $Q^2$ quarks}. On the other hand, in the dual theory with higgsed (i.e. not confined but screened) ${\ov q}^1$ and $q_1$ dual quarks with $SU(\no)$ dual colors and with such scale of masses, there are e.g. {\bf only fixed numbers of equal mass bosons with fixed quantum numbers}: $\no(2\nd-\no)$ massive dual gluons and the same number of their scalar superpartners.

{\bf Therefore, the mass spectra at this scale are clearly distinguishable in the direct and dual theories}.\\

3)  The next scale is $m_{Q,2}^{\rm pole}\sim \mu^{\rm pole}_{\dq,2}\sim\mu^{\rm pole}_3(\Phi_2^2)\sim\langle\lym\rangle\ll m_{Q,1}^{\rm pole}$, \eqref{(4.5)},\eqref{(4.9)},\eqref{(5.6)}. There are many gluonia in both direct and dual theories with such scale of masses and it seems these can be undistinguishable. Besides, there are e.g. many flavored quarkonia with different spins, with masses of this scale, made from confined quarks ${\ov Q}_2, Q^2$ quarks in the direct theory and from confined quarks $\odq^2, \dq_2$ in the dual one. These two sets of quarkonia can also be undistinguishable. But there are additionally $(\nt^2-1)$ elementary $SU(\nt)$ adjoint scalar superfields $\Phi^2_2$ with this scale of masses in the direct theory. And supposing that the number of scalar quarkonia $({\ov Q}_2 Q^2)$ and $(\odq^2\dq_2)$ is the same in the direct and dual theories, these extra $(\nt^2-1)$ elementary scalars $\Phi_2^2$ will distinguish these two theories.\\

4) And finally for particles with nonzero masses, there are $\no^2$ (i.e. $(\no^2-1)\,\,\, SU(\no)$ flavor adjoints plus one singlet) third generation lightest elementary scalar fields $(\Phi^{\rm pole}_3)_i^j,\, i,j=1...\no$ with $\mu^{\rm pole}_3(\Phi_1^1)\ll\langle\lym\rangle$ in the direct theory and the same number and the same (up to possible factors $O(1)$\,) mass  dual pions (nions) $N^i_j,\, i,j=1...\no$ in the dual one, \eqref{(4.8)},\eqref{(5.10)}. These two sets look undistinguishable (with our accuracy).\\

5) $2\no\nt$ fion fields $\Phi^1_2$ and $\Phi^2_1$ of the third generation in the direct theory, and the same number of nions (dual pions) $N^1_2$ and $N^2_1$ in the dual theory have the same quantum numbers and are all massless, they are Nambu-Goldstone particles of the spontaneously broken global flavor symmetry $U(N_F)\ra U(\no)\times U(\nt)$. These two sets are clearly undistinguishable.\\

{\bf On the whole, the mass spectra of the direct and dual theories in this region of the Lagrangian parameters are different (this is especially clearly seen in the point `2'), in disagreement with the Seiberg hypothesis about equivalence of such two theories}.\\

\hspace*{2cm} {\bf Part II\,:\,\,Mass spectra at    $0<\bd/N_F\ll 1\,,\,0<(2\no-\bd)/N_F\approx 2\no/N_F=O(1)$}\\

There is now the additional small parameter $0<\bd/N_F\ll 1,\, \bd=(3\nd-N_F)=(2N_F-3N_c)$, and this allows to see {\bf parametric differences} between mass spectra of the direct and dual theories.\\

At these values of parameters, the qualitative difference is that regimes at $\mu<m_{Q,1}^{\rm pole}$ are not conformal now. The direct theory is in the very strong coupling regime with $a(\mu\ll m_{Q,1}^{\rm pole})\gg 1$, while the dual theory at $\mu^{\rm pole}_{\dq,2} < \mu<{\ov\mu}^{\rm pole}_{\rm gl,1}$ is in the weakly coupled infrared free logarithmic regime. Not going into details, we note below only few qualitatively important points and give some results.

i) {\bf In the direct theory}. According to Seiberg's view of the standard direct (i.e. without fields $\Phi^i_j$)\, ${\cal N}=1$ SQCD at
$N_c+1<N_F<3N_c/2$, with the scale factor $\la$ of $SU(N_c)$ gauge coupling
\footnote{\,
and the same at $\mu=m_{Q,1}^{\rm pole}$ for the direct $\Phi$-theory considered here with $N_F\ra N^\prime_F=N_F-\no=\nt$ and $\la\ra
\Lambda^\prime=m_{Q,1}^{\rm pole}$
}
and direct quarks with $m_Q=0$ (or with $m_Q\ll\Lambda_Q$), the regime of the direct theory at $\mu<\la$ is in this case: {\bf `confinement without chiral symmetry breaking'} (as far as small $m_Q\neq 0$ can be neglected). And {\bf the dual theory is considered as the lower energy form of the direct theory}. This means that all direct quarks remained massless (or light), but hadrons made from these massless (or light) quarks and direct gluons {\bf acquired large masses $\sim\la$ due to mysterious confinement with the string tension} $\sigma^{1/2}\sim\la$, and decoupled at $\mu<\la$. Instead of them, there mysteriously appeared massless (or light) composite solitons. These last are particles of the dual theory.

This picture was questioned in \cite{ch1} (see section 7 therein). It was argued that, with the unbroken chiral flavor symmetry $SU(N_F)_L\times SU(N_F)_R$ and unbroken R-charge, it is impossible to write at $\mu\sim\la$ the nonsingular superpotential of the effective Lagrangian of massive flavored hadrons with masses $\sim\la$ made from direct massless (or light) quarks.~
\footnote{\,
This is similar to our ordinary QCD with confinement, with massless (or light) quarks but without chiral symmetry breaking. It is impossible then e.g. to have in the effective hadron Lagrangian at $\mu\sim\Lambda_{QCD}$ the massive nucleons with the mass $\sim\Lambda_{QCD}$, as the term $\sim\Lambda_{QCD} {\ov N} N$ in the potential is incompatible with the unbroken chiral symmetry. And the situation in ${\cal N}=1$ SQCD is even more restrictive because the superpotential is holomorphic and due to additional R-charge conservation.
}

We also recall here the following. There is no confinement in Yukawa-like theories without gauge interactions. The confinement originates {\bf only} from the unbroken YM, or ${\cal N}=1$ SYM in ${\cal N}=1$ SQCD-like theories. And because ${\cal N}=1$ SYM has only one dimensional parameter $\langle\lym\rangle=\langle S\rangle^{1/3}$, the string tension is $\sigma^{1/2}\sim\langle\lym\rangle$. But in the standard ${\cal N}=1$ SQCD the value of $\lym$ is well known: $\lym=(\Lambda_Q^\bo\det m_Q)^{1/3N_c}\ll\Lambda_Q$. Therefore, such SYM cannot produce confinement with the string tension $\sim\Lambda_Q$ (and there is no confinement at all at $m_Q\ra 0$).~
\footnote{\,
And the same for the direct SQCD-like $\Phi$-theory considered here:\, $\langle\lym\rangle=(\la^{\bo}\det\langle m^{\rm tot}_Q\rangle)^{1/3N_c}\ll\Lambda^\prime
=m_{Q,1}^{\rm pole}$. Therefore, such SYM cannot produce confinement with $\sigma^{1/2}\sim m_{Q,1}^{\rm pole}$, only with $\sigma^{1/2}\sim\langle\lym\rangle
\ll m_{Q,1}^{\rm pole}$.
}

For these reasons, we used below the picture described in section 7 of \cite{ch1}. I.e., in our case here with $\bd/N_F\ll 1$, after the direct quarks ${\ov Q}_1, Q^1$ decoupled as heavy at $\mu< m_{Q,1}^{\rm pole}$, the remained direct theory with light $SU(N_c)$ gluons and $\nt > N_c$ light quark flavors, $1<\nt/N_c < 3/2$, enters smoothly at lower energy into the perturbative (very) strong coupling regime with $a(\mu\ll m_{Q,1}^{\rm pole})\sim (m_{Q,1}^{\rm pole}/\mu)^{\nu=\frac{3N_c-2\nt}{\nt-N_c}}\gg 1$, and with all its colored particles effectively massless at $m_{Q,2}^{\rm pole}<\mu< m_{Q,1}^{\rm pole}$. (And NSVZ $\beta$-function \cite{NSVZa} allows this). The anomalous dimension of quarks ${\ov Q}_2, Q^2$ in the range $m_{Q,2}^{\rm pole}<\mu< m_{Q,1}^{\rm pole}$ is in this regime: $\,\gamma^{\prime}_{Q,2}=(2 N_c-\nt)/(\nt-N_c)>1$ \cite{ch1,ch3}, while those of $\Phi_2^2$ is $\gamma^{\prime}_{\Phi_2^2}= - 2 \gamma^{\prime}_{Q,2}$. At $\mu< m_{Q,2}^{\rm pole}\sim (\mph/\mI) m_{Q,1}^{\rm pole}\gg\langle\lym\rangle$ the quarks ${\ov Q}_2, Q^2$ decouple as heavy in the (very) strong coupling regime, and there remains ${\cal N}=1\,\, SU(N_c)$ SYM with its scale factor $\langle\lym\rangle$ determined from matching of couplings at $\mu=m_{Q,2}^{\rm pole}$\,:
\bbq
\hspace*{-6mm}  \Bigl [\, a_{+}(\mu=m_{Q,2}^{\rm pole})=\Bigl (\frac{m_{Q,1}^{\rm pole}}{m_{Q,2}^{\rm pole}}\Bigr )^{\nu=\frac{3N_c-2\nt}{\nt-N_c}}\,\Bigr ]=
\Bigl [\, a_{SYM}(\mu=m_{Q,2}^{\rm pole})=\Bigl (\frac{m_{Q,2}^{\rm pole}}{\langle\lym\rangle}\Bigr )^3\, \Bigr ]\,\ra
\eeq
\bbq
\ra\,\langle\lym\rangle^3=\la^3\Bigl (\frac{\mph}{\la}\Bigr )^{\frac{n_2}{n_2-N_c}}\Bigl (\frac{m_1}{\la}\Bigr )^{\frac{\nt-n_1}{n_2-N_c}}\,,
\eeq
as it should be, see \eqref{(4.7)}.\\

ii) {\bf In the dual theory}. This enters into the IR-free weakly coupled logarithmic regime at $\mu^{\rm pole}_{\dq,2} < \mu<{\ov\mu}^{\rm pole}_{\rm gl,1}$, and the dual quarks $\odq^2, \dq_2$ with $(\nd-\no)$ non-higgsed colors and $\nt > N_c$ flavors decouple as heavy at $\mu<\mu^{\rm pole}_{\dq,2}$. There remains ${\cal N}=1\,\, SU(\nd-\no)$ SYM with the same scale factor $\langle\lym\rangle\ll\mu^{\rm pole}_{\dq,2}$.

The parameter $Z_q$ of the dual theory is exponentially small at $\bd/N_F\ll 1$. Its value is determined from matching at $\mu=\mu^{\rm pole}_{\dq,2}$ of couplings ${\ov a}_{+}$ of higher energy  ${\cal N}=1$ SQCD with $SU(\nd-\no)$ colors and with quarks $\odq^2, \dq_2$ with $\nt$ flavors, and ${\ov a}_{-}$ of lower energy $SU(\nd-\no)$ ${\cal N}=1$ SYM (see \eqref{(6.3)},\eqref{(6.4)}\,)  \,:
\bbq
\Biggl [\frac{1}{{\ov a}_{+}}\approx \frac{1}{{\ov a}_{*}}+\frac{2\no-\bd}{\nd-\no}\log\Bigl (\frac{{\ov\mu}^{\rm pole}_{\rm gl,1}}{\mu^{\rm pole}_{\dq,2}}\Bigr )\Biggr ]=\Biggl [ \frac{1}{{\ov a}_{-}}\approx 3\log \Bigl (\frac{\mu^{\rm pole}_{\dq,2}}{\langle\lym\rangle}\Bigr )\Biggr ],\,\,\,
\frac{1}{{\ov a}_{*}}=\frac{N_F}{\bd}\,\, \ra \,\, Z_q\sim\exp\{-\frac{\nd-\no}{\bd}\}\ll 1.
\eeq

{\bf A)} {\bf Strongly coupled direct theory}

\vspace*{2mm}

a) All $N_F^2$ masses of second generation fions $\mu^{\rm pole}_2(\Phi_i^j)=\mu_o$ remain the same as before \eqref{(4.2)}.\\

b) The masses of $m_{Q,1}^{\rm pole}$ and $\mu^{\rm pole}_{3}(\Phi_1^1)$ are frozen at $\mu<m_{Q,1}^{\rm pole}$ and so remain
the same as before \eqref{(4.3)},\eqref{(4.8)} (but now, at $(2\no-\bd)>0,\,\, \mu^{\rm pole}_{3}(\Phi_1^1)\gg\langle\lym\rangle$ in \eqref{(4.8)}).\\

c) The mass of $m_{Q,2}^{\rm pole}$ looks now as:\, $\langle\lym\rangle\ll m_{Q,2}^{\rm pole}=(\mph/\mI) m_{Q,1}^{\rm pole}\ll m_{Q,1}^{\rm pole},\\
\mI=\la(\la/m_1)^{(2N_c-N_F)/N_c}$, compare with \eqref{(4.5)}.\\

d) The masses of $\nt^2$ fions $\mu^{\rm pole}_{3}(\Phi_2^2)$ are parametrically smaller now than before, they become the smallest masses among
all other nonzero masses, compare with \eqref{(4.9)}
\vspace*{-2mm}
\bq
\mu^{\rm pole}_3(\Phi_2^2)\sim \Bigl (\frac{\mph}{\mo}\Bigr )^{\frac{2(2\no-\bd)}{3(\nt-N_c)}\,>\,0}\langle\lym\rangle\ll \langle\lym\rangle\,.\label{(6.1)}
\eq

e) $2\no\nt$ fion fields $\Phi^1_2$ and $\Phi^2_1$ of the third generation are massless as in Part I above.\\

\vspace*{1.5mm}

{\bf B)} {\bf Weakly coupled dual theory},\,\, $(\nd-\no)/\bd\gg 1$
\vspace*{1mm}

For simplicity, we ignore logarithmic factors of the dual quark RG-evolution at $\mu^{\rm pole}_{\dq,2}<\mu<{\ov\mu}^{\rm pole}_{\rm gl,1}$.\\

a) All $N_F^2$ equal mass $\mu^{\rm pole}(M^i_j)$ mions of the dual theory and $N_F^2$ equal mass $\mu^{\rm pole}_2(\Phi_i^j)$ of second generation
fions in the direct theory have now parametrically different masses, compare with \eqref{(4.2)},\eqref{(5.2)},
\footnote{\,
Here and below we trace only factors which are the exponentially small (or large) in their dependence on the small parameter $\bd/N_F\ll 1$,
i.e.powers of $Z_q\sim\exp\{- (\nd-\no)/\bd\}\ll 1$, and ignore preexponential power-like in $\bd/N_F$ factors. Besides, $Z_q$ does not compete in any way
in its smallness with e.g. $m_Q/\la\ll 1$ or $\mph/\mo\ll 1$.
}
\vspace*{-4mm}
\bq
\mu^{\rm pole}(M^i_j)\sim Z^2_q\,\mu^{\rm pole}_2(\Phi_i^j)\ll \mu^{\rm pole}_2(\Phi_i^j)\,,\quad Z_q\sim\exp\{-\frac{\nd-\no}{\bd}\}\ll 1\,.\label{(6.2)}
\eq
\vspace*{-6mm}

\hspace*{3mm} b) ${\ov\mu}^{\rm pole}_{\rm gl,1}$ is parametrically smaller now than before, compare with \eqref{(4.3)},\eqref{(5.3)},
\vspace*{-3mm}
\bq
{\ov\mu}^{\rm pole}_{\rm gl,1}\sim Z^{1/2}_q\, m_{Q,1}^{\rm pole}\ll m_{Q,1}^{\rm pole}\,.\label{(6.3)}
\eq
\vspace*{-3mm}
c) $\mu^{\rm pole}_{\dq,2}$ looks now as, compare with \eqref{(4.5)},\eqref{(5.6)},

\vspace*{-1mm}
\bq
\mu^{\rm pole}_{\dq,2}\sim\frac{1}{Z_q}\Bigl (\frac{\mo}{\mph}\Bigr )^{\frac{2\no-\bd}{3(\nt-N_c)}\,>\,0}\langle\lym\rangle
\gg\langle\lym\rangle\,,\quad\mu^{\rm pole}_{\dq,2}\sim\frac{1}{Z_q}\,m_{Q,2}^{\rm pole}\gg m_{Q,2}^{\rm pole}\gg\langle\lym\rangle\,.\label{(6.4)}
\eq

\vspace*{-3mm}

\bbq
\mu^{\rm pole}_{\dq,2}\sim\Bigl (\frac{\mph}{Z^{3/2}_q\mo}\Bigr )\,{\ov\mu}^{\rm pole}_{\rm gl,1}\ll{\ov\mu}^{\rm pole}_{\rm gl,1}\,,\quad
\mph\ll Z^{3/2}_q\mo\,.
\eeq

\vspace*{-2mm}

Both direct quarks ${\ov Q}_2,\, Q^{2}$ and dual ones $\odq^2,\, \dq_2$ are weakly confined (i.e. the string tension originating from corresponding SYMs is parametrically smaller than quark masses, $\sigma^{1/2}\sim\langle\lym\rangle\ll m_{Q,2}^{\rm pole}\ll\mu^{\rm pole}_{\dq,2}$) and form a large number of various quarkonia. But quarks $\odq^2,\, \dq_2$ are non-relativistic and weakly coupled inside low lying quarkonia in the dual theory, so that the mass splittings between adjacent levels of dual quarkonia are parametrically small, $\delta m/m\sim O(\bd^{\,2}/N_F^2)\ll 1$, while there is nothing similar in the strongly coupled direct theory.

d) $\no^2$ fields $N_1^1$ of the dual theory and $\no^2$ fields $\Phi_1^1$ of the third generation fions of the direct theory, both sets with the same quantum numbers, also have now parametrically different masses, compare with \eqref{(4.8)},\eqref{(5.10)}, but now at $(2\no-\bd)>0$,
\vspace*{-3mm}
\bq
\mu^{\rm pole}(N_1^1)\sim \frac{1}{Z_q}\mu_{3}^{\rm pole}(\Phi^1_1)\gg \mu_{3}^{\rm pole}(\Phi^1_1)\gg\langle\lym\rangle\,.\label{(6.5)}
\eq
\vspace*{-3mm}

e) The low energy frozen "masses" of mions are also changed. $z_{M}^{\prime}$ factor in \eqref{(5.9)} is only logarithmic now (and is ignored). Therefore,  now instead of \eqref{(5.8)}:
\vspace*{-3mm}
\bq
\mu_{low}(M_i^j)\sim\Bigl (\frac{Z^{3/2}_q\mo}{\mph}\Bigr )\muo\gg\muo\,.\label{(6.6)}
\eq

\vspace*{-3mm}

f) $2\no\nt$ nion fields $N^1_2$ and $N^2_1$ (dual pions) of the dual theory are massless as in Part I above and are undistinguishable from the $2\no\nt$ third generation massless fion fields $\Phi^1_2$ and $\Phi^2_1$ of the direct theory. All these particles are Nambu-Goldstone particles of the spontaneously broken global symmetry $U(N_F)\ra U(\no)\times U(\nt)$.

It is seen that at the left end of the conformal window, i.e. at $\bd/N_F\ll 1$ in this Part II, {\bf in addition to clear qualitative differences in point `2'} of Part I above at $\bd/N_F=O(1)$, all corresponding nonzero mass scales of the direct and dual theories are now {\bf parametrically different} in this region of the Lagrangian parameters: they differ at least by powers of the {\bf parametric factor} $Z_q\sim\exp\{- (\nd-\no)/\bd \}\ll 1\,.$
\footnote{\,
And there are no particles now in the dual theory with the scale of masses similar to $\mu^{\rm pole}_3(\Phi_2^2)$ in \eqref{(6.1)}.
}
And logarithmic factors of the RG-evolution of $\odq^2,\, \dq_2$ quarks present in the dual theory (and absent in the direct one) result in additional parametric differences of corresponding masses.

Therefore, there are no reasons for these corresponding masses to become exactly equal at $\bd/N_F=O(1)$ in Part I above.\\

On the whole, we conclude that, {\bf although clearly surprisingly similar in a number of respects, the direct and Seiberg's dual ${\cal N}=1$ SQCD-like theories  have different mass spectra and are not equivalent}. As was shown above, this is  especially clearly seen at the left end of the conformal window at $0<\bd/N_F\ll 1$ considered here, where {\bf the corresponding mass scales are parametrically different}.

Recall that methods of mass spectra calculations used e.g. in \cite{ch2,ch3} and in all cases considered above satisfy all those tests which were used as checks of the Seiberg hypothesis about equivalence of the direct and dual theories. This shows that all those tests, {\bf although necessary, are not sufficient}. (And similarly at both ends of the conformal window, at the left end $(3\nd-N_F)/N_F\ll 1$, or at the right end $(3 N_c-N_F)/N_F\ll 1$ in the standard ${\cal N}=1$ SQCD and its Seiberg's dual, i.e. both without fields $\Phi$, see ~{\cite{ch3}}).

In addition, {\bf we see no any reasons to interpret the dual theory as "the low energy solitonic magnetic" form at $\mu<\la$ of the direct fundamental electric theory}. This is evident e.g. in the standard ${\cal N}=1$ SQCD within the conformal window, where the UV free direct theory enters smoothly at $\mu<\la$ into the perturbative conformal regime with all its quarks and gluons remaining effectively massless. And as was argued e.g. in \cite{ch1,ch3} and above in this section, at $N_c<N_F<3N_c/2$ also. The only difference is that the regime at $\mu<\la$ will be not conformal but very strong coupling one at $N_c<N_F<3N_c/2$.
\footnote{\,
In the case considered here this happens also at  $\mu<m_{Q,1}^{\rm pole}$ and  $0<\bd/N_F\ll 1\,,\,0<(2\no-\bd)/N_F\approx 2\no/N_F=O(1)$\,.
}
The "dual" theory has to be considered simply as {\bf a definite independent theory}. And both theories can be compared at $\mu<\la$ to see whether they are equivalent or not.

On the other hand, it seems clear that, indeed, {\bf there is some hidden symmetry} (broken by $m_Q\neq 0$ and, in our case here, by $\la\ll\mph\ll\mo=\la (\la/m_Q)^{(2N_c-N_F)/N_c}$) {\bf which makes direct and Seiberg's dual ${\cal N}=1$ SQCD-like theories, although not equivalent, but very similar}, see Appendix A. And described above methods of mass spectra calculation for such theories at (very) strong couplings demonstrate this. And, from
our viewpoint, just this is most important. This shows that we understand the dynamics of such theories sufficiently well.

Much more examples can be found in \cite{ch2}. See also \cite{ch3} and Appendix B about mass spectra in the standard SQCD and similar problems with its Seiberg's dual variant.
\footnote{\,
Recall that at $N_c< N_F <3 N_c/2$ the standard ${\cal N}=1$ SQCD with light quarks, $m_Q\ll\la$, and its Seiberg's dual (i.e. both without fields $\Phi$) also have qualitatively different mass spectra \cite{ch3}. In the strongly coupled at $\mu<\la$ direct theory the quark masses are $m^{\rm pole}_Q\sim \la (m_Q/\la)^{\nd/N_c},\,\,\nd=(N_F-N_c)$. These strongly coupled but weakly confined by strings with the tension $\sigma^{1/2}\sim\lym\ll\la$ quarks decouple at $\mu<m^{\rm pole}_Q$ and there remains ${\cal N}=1$ $SU(N_c)$ SYM with $\lym=(\la^\bo m^{N_F}_Q)^{1/3 N_c}$. Up to additional logarithmic factors, in the IR-free at $\mu<\la$ dual theory the weakly coupled and weakly confined dual quarks have masses $\mu^{\rm pole}_q\sim m^{\rm pole}_Q$. After they decouple, there remains dual ${\cal N}=1$ $SU(\nd)$ SYM with the same $\lym$ and $N_F^2$ lighter mions $M^i_j$ with masses $\mu^{\rm pole}(M^i_j)\sim \la (m_Q/\la)^{(N_c-\nd)/N_c}\ll\lym\ll\mu^{\rm pole}_q$. There is no analog of these light mions in the direct theory. \label{(f9)}
}

\vspace*{-6mm}

\appendix
\section{'t Hooft triangles,\,\, $0<\bd/N_F=O(1)\,,\,0<(\bd- 2\no)/N_F=O(1)$}
\vspace*{-3mm}

The quantum numbers of various fields with respect to the global $SU(N_F)_L\times SU(N_F)_R$ chiral symmetry are the following. -

a) The direct quarks are $Q^L$ and ${\ov Q}_{\ov R}$, i.e. $Q$ realizes the fundamental representation $N_F$ of $SU(N_F)_L$, while ${\ov Q}$ - the
antifundamental representation ${\ov N}_F$ of $SU(N_F)_R$.

b) The fions $\Phi^i_j$ are $\Phi_{\ov L}^R\,$.

c) The dual quarks are $q_{\ov L}$ and ${\ov q}^R$, while mions $M^i_j$ are $M^L_{\ov R}$ and the nions (dual pions) are $N^R_{\ov L}$.

It is worth noting also that 't Hooft triangles have different values at different ranges of scales because the chiral flavor symmetries are broken not spontaneously but explicitly by $m_Q\neq 0$ and $\la\ll\mph\ll\mo$.

{\bf 1)} The range $\mu_2^{\rm pole}(\Phi^i_j)\sim\mu_o\sim \mu^{\rm pole}(M^i_j)\ll\mu\ll\la$.

All particles of the direct and dual theories, except for $N_F^2$ fions $\Phi^i_j$, are relevant in this range. The triangle $SU^{3}(N_F)_L$ is $N_c$ in the direct theory, while in the dual one it is $(-{\ov N}_c)$ from dual quarks and $N_F$ from $N_F^2$ mions $M^i_j$, i.e. also $N_c$ on the whole. This case was checked by Seiberg in \cite{S2}.

{\bf 2)} The range $m_{Q,1}^{\rm pole}\sim \muo^{\rm pole}\ll\mu\ll \mu_2^{\rm pole}(\Phi^i_j)$

All particles of the direct and dual theories, except for $N_F^2$ mions $M^i_j$, are relevant in this range.  The triangle $SU^{3}(N_F)_L$ is $(-{\ov N}_c)$
in the dual theory, while in the direct one it is $N_c$ from direct quarks and $(- N_F)$ from $N_F^2$ fions $\Phi^i_j$, i.e. also $(-{\ov N}_c)$ on the whole.

{\bf 3)} $m_{Q,2}^{\rm pole}\sim \mu^{\rm pole}_{\dq,2}\sim\mu^{\rm pole}_3(\Phi_2^2)\sim\langle\lym\rangle \ll \mu\ll m_{Q,1}^{\rm pole}$

a) The triangle $SU^{3}(\no)_L$. The nions $N^1_1$ give $(-\no)$ and nions $N_1^2$ give $(-\nt)$ in the dual theory, i.e. $ (- N_F)$ on the whole.
In the direct theory:  fions $\Phi^1_1$ give $(-\no)$ and fions $\Phi^2_1$ give $(-\nt)$, i.e. also $(- N_F)$ on the whole.

b)  The triangle $SU^{3}(\nt)_L$. In the dual theory: quarks $\dq_2$ give $(-\nd+\no)$ and nions $N^1_2$ give $(-\no)$, i.e. $(-\nd)$ on the whole.
In the direct theory:  quarks $Q^2$ give $N_c$, fions $\Phi^1_2$ give $(-\no)$ and fions $\Phi^2_2$ give $(-\nt)$, i.e. also $(-\nd)$ on the whole.

{\bf 4)} The range $\mu_{3}^{\rm pole}(\Phi^1_1)\sim\mu^{\rm pole}(N_1^1) \ll\mu \ll \langle\lym\rangle$

a) The triangle $SU^{3}(\no)_L$. In the dual theory: nions $N^1_1$ give $(-\no)$ and $N^2_1$ give $(-\nt)$, i.e. $(-N_F)$ on the whole.
In the direct theory: fions $\Phi^1_1$ give $(-\no)$ and $\Phi^2_1$ give $(-\nt)$, i.e. also $(-N_F)$ on the whole.

b) The triangle $SU^{3}(\nt)_L$. In the dual theory: nions $N^1_2$ give $(-\no)$, while in the direct theory fions $\Phi^1_2$ also give $(-\no)$.

{\bf 5)} The range $0\leq\mu\ll \mu_{3}^{\rm pole}(\Phi^1_1)$.

a) The triangle $SU^{3}(\no)_L$. In the direct theory: the fions $\Phi_1^2$ give $(-\nt)$. In the dual theory: nions $N^2_1$ also give $(-\nt)$.

b) The triangle $SU^{3}(\nt)_L$. In the direct theory: the fions $\Phi^1_2$ give $(-\no)$. In the dual theory: nions $N^1_2$ also give $(-\no)$.

It is seen that triangles are the same in this case, while there are differences in the mass spectra (see Part I Conclusions above). This shows once more that the equality of triangles, although necessary, is not sufficient to speak about the equivalence of two theories.

At $\bd/N_F\ll 1$ in Part II Conclusions above, there appear parametric intervals of scales within which the corresponding masses of the direct and dual theories are different. And within these intervals the triangles of the direct and dual theories are also different.\\

\vspace*{-10mm}

\section{Mass spectra in the standard ${\cal N}=1$ SQCD and its Seiberg's dual at $N_F=N_c+1$.}

\hspace*{ 4mm} According to \cite{ch3} (see also the arguments in Part II of Conclusions and the footnote \ref{(f9)}), the mass spectrum of the direct theory is a smooth continuation to $N_F=N_c+1$ of those at $N_c+1<N_F<3 N_c/2$. I.e., the quark masses are $m_Q^{\rm pole}=C_Q\la (m_Q/\la)^{1/N_c}\ll\la,\,\,C_Q={\cal O}(1)$. After they decouple at $\mu<m_Q^{\rm pole}$, there remains ${\cal N}=1\,\,SU(N_c)$ SYM with its scale factor $\lym=\la (m_Q/\la)^{(N_c+1)/3 N_c}\ll m_Q^{\rm pole}$. This is all. The $SU(N_c)$ gluonia with the mass scale $\sim\lym$ are the lightest particles.

Now, about Seiberg's dual theory. It was proposed by Seiberg in \cite{S1} that all $SU(N_c)$ gluons and all light quarks  with masses $m_Q=m_Q(\mu=\la)\ll\la$ and $N_F=N_c+1$ flavors of the direct theory are confined by strings with the strong tension $\sigma^{1/2}\sim\la$ and form hadrons with masses $\sim\la$.
And all this hadrons decouple as heavy at scales $\mu < \la$. And he proposed that, instead of them, there appear light colorless solitons (dual particles): mesons $M^i_j$ and baryons $B_i,\, {\ov B}^j,\,\,i,j=1...N_F=N_c+1$.  In \cite{IS} this regime (at $m_Q\ra 0)$ with $N_F=N_c+1$ was called  as ''confinement without the chiral symmetry breaking''\,.

After decoupling of all heavy hadrons, the proposed in \cite{S1} Lagrangian of these light mesons and baryons  has the form at $\mu=\la$
\bq
K_{\rm dual}={\rm Tr}_{N_c+1}\,\frac{M^\dagger M}{\la^2}+{\rm Tr}_{N_c+1}\,(B^\dagger B+{\ov B}^\dagger {\ov B})\,,\label{(B1)}
\eq

\vspace*{-4mm}

\bbq
{\cal W}_{\rm dual}=m_Q{\rm Tr}_{N_c+1} (M)+{\rm Tr}_{N_c+1}\,({\ov B}\frac{M}{\la} B)-\frac{\det_{N_c+1} M}{\la^{2N_c-1}}\,.
\eeq

It was pointed out in the subsequent paper \cite{S2} that the Lagrangian \eqref{(B1)} of the dual theory with $N_F=N_c+1$ can be obtained e.g. as follows. One can start with the direct $SU(N_c)$ theory with $N_F^\prime=N_c+2$ flavors of direct quarks and with the Lagrangian at $\mu=\la$
\bq
K_{direct}={\rm Tr}_{N_c+2}\, (Q^\dagger Q+Q\ra {\ov Q})\,,\,\quad {\cal W}_{direct}={\cal W}_{gauge}(SU(N_c))+m_Q {\rm Tr}_{N_c+1}\,({\ov Q} Q)+
\la ({\ov Q}_{0} Q^{0})\,.\label{(B2)}
\eq
Integrating out at $\mu<\la$ the last heavy quarks with the mass $\la$, one obtains the desired direct $SU(N_c)$ theory with $N_F=N_c+1$ light flavors of direct quarks $Q^i,\, i=1...N_c+1$, with masses  $m_Q\ll\la$.

On the other hand, the proposed in \cite{S2} Lagrangian of the dual $SU(\nd=2)$ theory with $N_F^\prime=N_c+2$ dual quark flavors looks at $\mu=\la$ as

\vspace*{-8mm}

\bq
K_{\rm dual}={\rm Tr}_{N_c+2}\frac{M^\dagger M}{\la^2}+ {\rm Tr}_{N_c+2}\Bigl (\, q^\dagger q + (q\ra {\ov q}\, \Bigr )\,,
\eq

\vspace*{-4mm}

\bbq
{\cal W}_{dual}={\cal W}_{gauge}(SU(2))+m_Q {\rm Tr}_{N_c+1}( M) +\la M^{0}_{0} - {\rm Tr}_{N_c+2}\Bigl (\,{\ov q}\frac{M}{\la} q\,\Bigr )\,.\label{(B3)}
\eeq

From \eqref{(B2)} and Konishi anomalies \cite{Konishi}, the mean vacuum values of mions $M^i_j\ra ({\ov Q}_j Q^i)$ and dual quarks look at $\mu=\la$  as

\vspace*{-4mm}

\bq
\langle M^i_j\rangle=\delta^i_j \frac{\langle S\rangle}{m_Q}=\delta^i_j \la^2\Bigl (\frac{m_Q}{\la}\Bigr )^{\frac{1}{N_c}}\,,\quad \langle M^{0}_{0}\rangle=\frac{\langle S\rangle}{\la}=\la^2 \Bigl (\frac{m_Q}{\la}\Bigr )^{\frac{N_c+1}{N_c}}\,, \quad \langle S\rangle=\la^3\Bigl (\frac{m_Q}{\la}\Bigr )^{\frac{N_c+1}{N_c}}\,,\label{(B4)}
\eq

\vspace*{-3mm}

\bbq
\langle {\ov q}^{0} q_{0}\rangle=\frac{\langle S\rangle \la}{\langle M^{0}_{0}\rangle}=\la^2\,,\quad \langle {\ov q}^j q_i\rangle=\frac{\langle S\rangle \la}{\langle M^{i}_{j}\rangle}=\delta_i^j m_Q\la\,, \quad i,j=1...N_c+1\,.
\eeq

It is seen from \eqref{(B4)} that the condensate  $\langle {\ov q}^{0} q_{0}\rangle^{1/2}$ of last dual quarks is much larger than their mass $\langle M^0_0\rangle/\la$. Therefore, they are higgsed and broke the whole dual $SU(\nd=2)$ group at the scale $\sim\la$. After integrating all heavy particles of the dual theory with masses $\sim\la$, there remains the IR-free dual theory with $N_F^2$ light mions $M^i_j$ and $N_F=N_c+1$ light quarks $q_i,\,{\ov q}^j$ with one screened color. Let us reassign them as $B_i=i\epsilon_{\alpha\beta}\langle q_{0}^{\alpha}\rangle q_i^{\beta}/\la,\,\,\,{\ov B}^j=i\epsilon^{\alpha\beta}\langle {\ov q}^{0}_{\alpha}\rangle{\ov q}^j_{\beta}/\la,\,\,\,\alpha,\beta=1,2$. In the range of scales $\mu^{\rm pole}(B)<\mu<\la$ all these particles are effectively massless and the Lagrangian looks as
\footnote{\,
We account in this range of scales of the IR-free theory \eqref{(B5)} for the parametric logarithmic renormalization factors $z_M$ and $z_B$ of $M$ and $B$.
}
(compare with \eqref{(B1)})

\vspace*{-5mm}

\bbq
K_{\rm dual}=z_M(\la,\mu){\rm Tr}_{N_c+1}\frac{M^\dagger M}{\la^2}+ z_B(\la,\mu){\rm Tr}_{N_c+1}\Bigl (\, B^\dagger B + (B\ra {\ov B}\, \Bigr )\,,
\eeq

\vspace*{-5mm}

\bq
{\cal W}_{\rm dual}=m_Q {\rm Tr}_{N_c+1} M +{\rm Tr}_{N_c+1}\Bigl (\,{\ov B}\frac{M}{\la} B\Bigr )-\int d{\tilde\mu}\,\frac{\det_{N_c+1}({\ov B} B)}{\la^{2N_c-1}}\,,\label{(B5)}
\eq
where $d{\tilde\mu}$ is the corresponding measure. The term $\sim\det_{N_c+1}({\ov B} B)$ in \eqref{(B5)} is a shorthand for the instanton contribution in the form of the multiquark 't Hooft operator. At $\mu<\mu^{\rm pole}(B)$ the baryons decouple as heavy and the Lagrangian looks as (compare with \eqref{(B1)})

\vspace*{-4mm}

\bbq
K_{\rm dual}=z_M(\la,\mu^{\rm pole}(B)){\rm Tr}_{N_c+1}\frac{M^\dagger M}{\la^2}+z_B(\la,\mu^{\rm pole}(B)){\rm Tr}_{N_c+1}\Bigl (\, B^\dagger B +
(B\ra {\ov B}\, \Bigr )\,,
\eeq

\vspace*{-3mm}

\bq
{\cal W}_{\rm dual}=m_Q {\rm Tr}_{N_c+1}( M)+{\rm Tr}_{N_c+1}\Bigl (\,{\ov B}\frac{M}{\la} B\Bigr ) -\frac{\det_{N_c+1}(M)}{\la^{2N_c-1}}\,.\label{(B6)}
\eq

\vspace*{-6mm}

The pole masses of baryons $B_i,\,{\ov B}^j$ and mions $M^i_j$ from \eqref{(B6)} are (compare with the footnote \ref{(f9)} at $\nd=N_F-N_c=1$)

\vspace*{-8mm}

\bq
\mu^{\rm pole}(B,{\ov B})=\frac{\la}{z_B(\la,\mu^{\rm pole}(B))}\Bigl (\frac{m_Q}{\la}\Bigr )^{\frac{1}{N_c}}\ll\la,\,\,
\mu^{\rm pole}(M)=\frac{\la}{z_M(\la,\mu^{\rm pole}(B))}\Bigl (\frac{m_Q}{\la}\Bigr )^{\frac{N_c-1}{N_c}}\ll\lym, \,\,\label{(B7)}
\eq

\vspace*{-3mm}

It is seen from \eqref{(B7)} that the mass spectrum at $\mu<\la$ of the dual theory with $N_F=N_c+1$ flavors is also its smooth continuation from $N_c+1<N_F<3N_c/2$ to $\nd=N_F-N_c=1$, see the footnote \ref{(f9)}, and only heavy dual gluons are absent, while the baryons $B_i, {\ov B}^j$ are really the remained light dual quarks.

On the whole, the mass spectra of the direct and dual theories with $N_F=N_c+1$flavors look as follows (see \cite{ch3}, Part II of Conclusions and the footnote \ref{(f9)} for the mass spectra of the direct theory).\\
1) In the direct theory. - There is a number of flavored hadrons with the mass scale $\sim m_Q^{\rm pole}\sim\la (m_Q/\la)^{1/N_c} \\
\ll\la$, with different spins and other quantum numbers. In addition, there is only a number of gluonia with the mass scale $\sim\lym=\la (m_Q/\la)^{(N_c+1)/3 N_c}\ll m_Q^{\rm pole}$. The real string tension in the direct theory originates from $SU(N_c)$ SYM and is not $\sigma^{1/2}\sim\la$ but $\sigma^{1/2}\sim\lym\ll\la$.\\
2) In the dual theory. - There are $2N_F$ baryons $B_i$ and ${\ov B}^j$ (= dual quarks). Their masses are $\mu^{\rm pole}(B)=
\Bigl (\la/z_B(\la,\mu^{\rm pole}(B))\Bigr )\Bigl  (m_Q/\la\Bigr )^{1/N_c}$. And there are $N_F^2$ lightest mions $M^i_j$ with masses $\mu^{\rm pole}(M)= \\
\Bigl (\la/z_M(\la,\mu^{\rm pole}(B))\Bigr )\Bigl (m_Q/\la\Bigr )^{(N_c-1)/N_c}\ll\lym \ll\mu^{\rm pole}(B)$. It is seen that the mass spectra of the direct
and dual are qualitatively different.

As for the 't Hooft triangles. - \\
1)\, In the range of scales (ignoring logarithmic factor $z_B$): $m_Q^{\rm pole}\sim\mu^{\rm pole}(B)\ll\mu\ll\la$. All triangles of the direct and dual theories are the same \cite{GW},\cite{V},\cite{S1}. \\
2)\, In the range $\lym\ll\mu\ll m_Q^{\rm pole}$. -  To the triangles $R^3$ and $R$ contribute $SU(N_c)$ gluinos of the direct theory and fermionic partners $\psi^i_j$  of mions $M^i_j$ in the dual one. The triangles in two theories are different. There are no contributions to other triangles in the direct theory, while there are contributions from $\psi^i_j$ to the $SU(N_F)^3_L$ and $SU(N_F)^2_L\times U(1)_R$ triangles.\\
3)\,  In the range of scales $\mu^{\rm pole}(M)\ll\mu\ll\lym$. - There are no contributions to all triangles in the direct theory, while still there are contributions from $\psi^i_j$ to the $SU(N_F)^3_L,\,\,SU(N_F)^2_L\times U(1)_R$, $R^3$ and $R$ triangles.

\end{document}